\documentstyle[12pt,epsfig]{article} 
\begin{document}
\title{Decoherence and bare mass induced by nonconformal metric fluctuations}
\author{ A. Camacho
\thanks{email: acamacho@nuclear.inin.mx} \\
Department of Physics, \\
Instituto Nacional de Investigaciones Nucleares\\
Apartado Postal 18--1027, M\'exico, D. F., M\'exico.}
\date{}
\maketitle

\begin{abstract}
The effects, upon the Klein--Gordon field, of nonconformal stochastic me\-tric fluctuations, are analyzed. It will be shown that these fluctuations allow us to consider an effective mass, i.e., the mass detected in a laboratory is not the parameter appearing in the Klein--Gordon equation, but a function of this parameter and of the fluctuations of the metric. In other words, in analogy to the case of a nonrelativistic electron in interaction with a quantized electromagnetic field, we may speak of a bare mass, where the observed mass shows a dependence upon the stochastic terms included in the metric.  Afterwards, we prove, resorting to the influence functional, that the energy--momentum tensor of the Klein--Gordon field inherites this stochastic behavior, and that this feature provokes decoherence upon a particle immersed in the region where this tensor is present. 

\end{abstract}
\bigskip
\bigskip


\section{Introduction}

The lack of a quantum theory of gravity has recently spurred the quest of possible detectable effects stemming from the different proposals that, currently, in this context exist. For instance, among these attempts we may mention the search for additional noise sources in gravity--wave interferometers [1], or the quest for modifications in the kinematical properties of light (a deformed dispersion relation) [2]. 

Here it will be assumed that quantum gravity corrections may be described as nonconformal stochastic fluctuations of the metric. The consequences of this assumption, in connection with some of the kinematical properties of light, have already been a\-nalyzed in a previous work [3], and now we extend this study and consider the possible effects of these fluctuations upon a particle satisfying the Klein--Gordon equation. Taking into account the presence of this stochasticity the canonical quantization of our particle will be carried out, and afterwards the energy--momentum tensor of the system will be calculated. The expectation value of the vacuum state will allow us to introduce, in analogy to the case of a nonrelativistic electron in interaction with a quantized electromagnetic field [4], the idea of bare mass, where in our case the observed mass shows a dependence on the stochastic terms included in the metric. 

Finally, resorting to the influence functional of Feynman and Vernon [5], it will be shown that the mass fluctuations of the Klein--Gordon field induce decoherence in the case of a particle located near our system. This fact is no surprise at all, since we already know that conformal fluctuations of the metric do render decoherence [6], though in the present work decoherence has a quite different origin, as will be shown below.

The present approach, though it comprises the possibility of having stochasti\-city in the mathematical model related to the description of spacetime, has not the same spirit as some previous works [7], which also include this feature as an important element to be considered. The main difference in connection with these results [7] involves the fact that they analyze the consequences, in the realm of semiclassical theo\-ry of gravity, of classical stochastic fluctuations in the spacetime geometry ste\-mming from quantum fluctuations of matter fields, whereas in our work these stochastic fluctuations are induced by quantum gravity effects, and then we study the consequences upon the quantum fluctuations of matter of them, and consider the emergence of decoherence as an {\it indirect} effect of quantum gravity. 
\bigskip
\bigskip

\section{Nonconformal metric fluctuations and Klein--Gordon equation}
\bigskip

At this point we assume that the spacetime metric undergoes nonconformal stochastic fluctuations, and that these fluctuations represent white noise. As mentioned in the first section, the introduction of this kind of behavior has already, though in a partial manner, been analyzed [3]. The nonconformal character now present implies that, from the outset, the maximal symmetry of the vacuum of the classical gravitational field will not be preserved. 

Hence, in the case where the average background metric is the Minkowkian one, we may write

{\setlength\arraycolsep{2pt}\begin{eqnarray}
ds^2 = e^{\psi(x)}\eta_{00}dt^2 + e^{\zeta(x)}\eta_{ij}dx^idx^j.
\end{eqnarray}}
\bigskip

Here we demand the following properties 

{\setlength\arraycolsep{2pt}\begin{eqnarray}
<e^{\psi(x)}\eta_{00}> = \eta_{00}, 
\end{eqnarray}}
 
{\setlength\arraycolsep{2pt}\begin{eqnarray}
<e^{\zeta(x)}\eta_{ij}> = \eta_{ij}. 
\end{eqnarray}}

From these last two expressions we conclude (assuming $\vert\psi(x)\vert<<1$) 

{\setlength\arraycolsep{2pt}\begin{eqnarray}
<\psi(x)> = 0, 
\end{eqnarray}}

{\setlength\arraycolsep{2pt}\begin{eqnarray}
<\partial_{\mu}\psi(x)> = 0. 
\end{eqnarray}}

These last conditions are related to the fact that $\psi(x)$ is white noise [3, 5]. Of course, $\zeta(x)$ fulfills the same conditions. From (5) we find that $<\psi(x)^2> = cte.$, and if these fluctuations have a gaussian behavior, then

{\setlength\arraycolsep{2pt}\begin{eqnarray}
<\psi^2(x)> = \sigma_1^2, 
\end{eqnarray}}

{\setlength\arraycolsep{2pt}\begin{eqnarray}
<\zeta^2(x)> = \sigma_2^2, 
\end{eqnarray}}

\noindent in these two last expressions $\sigma_1^2$ and $\sigma_2^2$ denote the corresponding square deviations.
\bigskip

The Lagrangian density for the case of a Klein--Gordon particle is [8]

{\setlength\arraycolsep{2pt}\begin{eqnarray}
{\mathcal{L}} = {\hbar^2\over 2m}\Bigl[-g(x)\Bigr]^{1/2}\Bigl(g^{\mu\nu}\phi_{,\mu}\phi_{,\nu} - [m^2c^2/\hbar^2 + \alpha R(x)]\phi^2\Bigr).
\end{eqnarray}}

Here $R(x)$ denotes the Ricci scalar and $\alpha$ is a numerical factor (defining the coupling between the scalar particle and the gravitational field) that in our case will be set equal to zero, i.e., we have the so called minimally coupled case [8]. Then the motion equation reads

{\setlength\arraycolsep{2pt}\begin{eqnarray}
{1\over c^2}{\partial^2\phi\over\partial t^2} + {2\over c^2}{\partial\phi\over\partial t}{\partial\psi\over\partial t} 
+ {3\over c^2}{\partial\phi\over\partial t}{\partial\zeta\over\partial t} + {m^2c^2\over\hbar^2}\phi  \nonumber\\
- e^{\phi- \zeta}\Bigl[\nabla^2\phi + \nabla\phi\cdot\nabla\psi + 4\nabla\phi\cdot\nabla\zeta\Bigr] = 0.
\end{eqnarray}}

It is readily seen (though it requires a messy calculation) that the solution to this last equation may be written in the form

{\setlength\arraycolsep{2pt}\begin{eqnarray}
\phi = <\phi>f(<\phi>, \psi, \zeta),
\end{eqnarray}}

where $<\phi>$ is the solution to case without stochastic terms in the metric ($\psi =0$ and $\zeta =0$), and $f$ is a function of $\psi$, $\zeta$, and $<\phi>$ .

Let us now choose periodic boundary conditions, namely, we confine the system to a cubic box with edge length $L$. This means

{\setlength\arraycolsep{2pt}\begin{eqnarray}
<\phi>_{n(\pm)} = \sqrt{mc^2\over L^3E_n}\exp\Bigl\{{i\over\hbar}(\vec{p}_n\cdot\vec{x} \mp E_nt)\Bigr\}.
\end{eqnarray}}

Clearly the boundary conditions imposed imply that $\vec{p}_n = {2\pi\hbar\over L}\vec{n}$, where $\vec{n}= (n_x, n_y, n_z)$, being $n_x, n_y$, and $n_z$ natural numbers, and $E_n = c\sqrt{m^2c^2 + p_n^2}$.

The general case becomes then

{\setlength\arraycolsep{2pt}\begin{eqnarray}
\phi(x,t) = \sum_n[c_n\phi_n + c_n^{\ast}\phi_n^{\ast}].
\end{eqnarray}}

At this point we consider the canonical quantization of the system, a fact that implies the introduction of the annihilation and creation operators ($a_n$ and $a^{\dagger}_n$, respectively), through the substitutions $c_n\rightarrow a_n$, and $c_n^{\ast}\rightarrow a^{\dagger}_n$ [4, 8].

Hence the quantized field reads 

{\setlength\arraycolsep{2pt}\begin{eqnarray}
\phi(x,t) = \sum_n[a_n<\phi>_nf(<\phi>_n, \psi, \zeta) + a^{\dagger}_n\Bigl(<\phi>_nf(<\phi>_n, \psi, \zeta)\Bigr)^{\ast}].
\end{eqnarray}}

Clearly this last expression shows that the stochasticity introduced in the metric tensor impinges upon the quantized field, and in consequence this characteristic must appear also in its energy--momentum tensor.

Proceeding in the usual way [8], the expectation value of the vacuum state of the time--time component of the energy--momentum tensor, $T_0^0$, has the following form 

{\setlength\arraycolsep{2pt}\begin{eqnarray}
<0\vert T_0^0\vert 0> = {mc^2\over L^3}\Bigl[\vert f_0\vert^2e^{3\zeta/2}\cosh(\psi/2) \nonumber\\
+ {\hbar^2\over 2m^2c^4}\vert{\partial f_0\over\partial t}\vert^2e^{(3\zeta - \psi)/2} + {\hbar^2\over 2m^2c^2}\vert\nabla f_0\vert^2e^{(\zeta + \psi)/2}\Bigr].
\end{eqnarray}}
\bigskip
\bigskip

\section{Conclusions}
\bigskip

Expression (14) allows us to state that the effective mass, associated to the quantized field, is not the parameter considered in the Lagrangian density, but

{\setlength\arraycolsep{2pt}\begin{eqnarray}
\tilde m = {m\over L^3}\int\Bigl[\vert f_0\vert^2e^{3\zeta/2}\cosh(\psi/2) \nonumber\\
+ {\hbar^2\over 2m^2c^4}\vert{\partial f_0\over\partial t}\vert^2e^{(3\zeta - \psi)/2} + {\hbar^2\over 2m^2c^2}\vert\nabla f_0\vert^2e^{(\zeta + \psi)/2}\Bigr]d^3x.
\end{eqnarray}}
\bigskip

Hence, we may, in analogy to the case of a nonrelativistic electron in interaction with a quantized electromagnetic field [4], assert that $m$ denotes the bare mass, and that the observed mass is $\tilde m$, and that it shows a stochastic behavior, inherited from the features of the metric. 

Going back to expression (14), and employing the statistical properties of our fields (expressions (4)-(7)) we find (after integrating) that 

{\setlength\arraycolsep{2pt}\begin{eqnarray}
\tilde m\approx m\Bigl[1 + {1\over 8}(\sigma_1^2 + 9\sigma_2^2)\Bigr].
\end{eqnarray}}

Let $\tilde L$ denote the largest distance between two points, such that they behave in a coherent way under the fluctuations $\zeta(x)$ (while $T$ is the corresponding time associated with $\psi(x)$). If we accept that: (i) these fluctuations are quantum gravity corrections to the Minkowskian metric, and (ii) Planck length might appear together with other length scales in the problem [9]; then we may introduce the following assumption

{\setlength\arraycolsep{2pt}\begin{eqnarray}
\sigma_1^2 = a^2L^2_p/\tilde L^2,  
\end{eqnarray}}

{\setlength\arraycolsep{2pt}\begin{eqnarray}
\sigma_2^2 = e^2T^2_p/T^2.  
\end{eqnarray}}

Here $T_P$ and $L_p$ denote the Planck length and time, respectively, while $a$ and $e$ are real numbers. This kind of relation, between square deviation and Planck length, has already been derived [10, 11]. Of course, this Ansatz requires a deeper analysis, and it does not discard other possibilities [12]. Clearly our model contains four free parameter, i.e., $e, a, T$, and $\tilde L$, which can not be deduced in the context of our assumptions.

Introducing the real number $\Gamma^2 = {1\over 8}\Bigl[a^2 + {9e^2\over\chi^2c^2 }\Bigr]$ (here we have defined $T = \chi\tilde L$) we may rewrite (16) as

{\setlength\arraycolsep{2pt}\begin{eqnarray}
\tilde m\approx m\Bigl[1 + \Gamma^2{L_p^2\over\tilde L^2}\Bigr].
\end{eqnarray}

In other words, the stochastic terms included in the metric tensor imply that the measurable mass is not the same as the mass parameter considered in the initial Lagrangian density, and that the difference between these two masses is a function of the largest distance at which two points do behave coherently under the action of the introduced metric fluctuations. In other words, this stochasticity characteristic does impinge upon the detectable properties of our system. 

An additional question that in this context could be addressed comprises the possible decoherence stemming from (14). In order to analyze this situation let us consider the Newtonian gravitational potential generated by (14) at a certain point with position vector $\vec{r}$ (clearly outside the cubic box employed above in the canonical quantization of the Klein--Gordon particle). 

{\setlength\arraycolsep{2pt}\begin{eqnarray}
V(r, t) = -{Gm\over r}\Bigl[1 + \hat V(t)\Bigr].  
\end{eqnarray}}

Here $\hat V(t)$ contains the stochastic terms stemming from (14). At this point the introduction of a Newtonian gravitational potential allows us to reinterpret the stochasticity in such a way that now, once again, we have distance fluctuations, namely $<r(t)>^{-1} = r^{-1}\Bigl[1 + \Gamma^2{L_p^2\over\tilde L^2}\Bigr]$, where $r$ on the right--hand side of this last expression is the parameter appearing in (20).
Consider now a quantum particle, whose mass is $M$, immersed in the aforementioned potential. According to the influence functional of Feynman and Vernon [5], the density matrix of $M$ reads 

{\setlength\arraycolsep{2pt}\begin{eqnarray}
\rho(\vec{r}_b, \vec{r}_c, t_f) = \int d\vec{r}_a d\vec{r}_e{\mathcal{D}}\vec{R}_a{\mathcal{D}}\vec{R}_e
\exp\Bigl\{{i\over\hbar}\Bigl[S(\vec{R}_a(t)) - S(\vec{R}_e(t)\Bigr]\Bigr\}\nonumber\\
\times F(\vec{R}_a(t), \vec{R}_e(t))
\rho(\vec{r}_a, \vec{r}_e, t_i).  
\end{eqnarray}}

Where the influence functional has been defined as [5]

{\setlength\arraycolsep{2pt}\begin{eqnarray}
F(\vec{R}_a(t), \vec{R}_e(t)) = \exp\Bigl\{-{i\over\hbar}\int_{t_i}^{t_f}\Bigl[\tilde V(\vec{R}_a(t), t) - \tilde V(\vec{R}_e(t)t)\Bigr]dt\Bigr\},
\end{eqnarray}}

with $\tilde V(\vec{r}(t), t) = -{Gm\over r(t)}\hat V(t)$. Hence (assuming $\Gamma^2\approx 1$, clearly this last parameter can not be predicted by our model, but the conclusions are not modified by this condition, and only for convenience we have imposed it)

{\setlength\arraycolsep{2pt}\begin{eqnarray}
{\partial\rho(\vec{r}_b, \vec{r}_c, t)\over\partial t} = {\mathcal{L}}(\rho) - 
\left\{\begin{array}{cl}
{2Gm^2\over\hbar\tilde L^3}\vert\vec{r}_b-\vec{r}_c\vert^2\rho(\vec{r}_b, \vec{r}_c, t), & \mbox{if $\vert\vec{r}_b-\vec{r}_c\vert\leq\tilde L$} \\
{2Gm^2\over\hbar\tilde L}\rho(\vec{r}_b, \vec{r}_c, t), & \mbox{if $\vert\vec{r}_b-\vec{r}_c\vert >\tilde L$}
\end{array}\right.
\end{eqnarray}}
\bigskip

The term ${\mathcal{L}}(\rho)$ denotes the unitary term involved in the time evolution of the density matrix, the one appears in the usual quantum theory [13], whereas the second one clearly implies the breakdown of unitary evolution, and provides the decay of the off--diagonal matrix elements [14].

In other words, even if we neglect the effects of the metric fluctuations (expre\-ssion (1)) upon the quantum particle, with mass $M$, the inherited stochasticity of the energy--momentum tensor, of the Klein--Gordon field, suffices to induce decoherence on $M$. As mentioned before, this is no surprise at all, since in the context of conformal metric fluctuations decoherence is also possible [6].

At this point it is noteworthy to add a comment concerning the manner in which the canonical quantization of the Klein--Gordon particle has been carried out. The e\-ffects of quantum gravity have been constrained to the modifications that the functions $f$ (expression (10)) imply, nevertheless the quantization procedure remains untouched. 
This last point has to be carefully studied. Indeed, in the context of the possible consequences of quantum gravity we may find the modification of the uncertainty principle [15]. This result that can be derived in the context of quantum geometry [16], black--hole effects [17], quantum measurements at Planck scale [18], in Newtonian gravity theory [19], or even resorting to a relation between the mass and the radius of a Schwarzschild black--hole [20]. 

This last remark leads us to the following question. If we take a look at the usual canonical quantization procedure we may see that in it the annihilation and creation operators ($a_n$ and $a^{\dagger}_n$, respectively) are defined resorting to two conditions (i) they are linear combinations of the position and momentum operator (see equations (2.24) of [4]), and (ii) the Heisenberg algebra between position and momentum operators suffers no changes. But if quantum gravity renders the modification of the usual uncertainty principle, does the definition of the creation and annihilation operators continue valid? 

Summing up, it has been shown that the introduction of nonconformal stochastic metric fluctuations renders the emergence of a bare mass, in the context of a Klein--Gordon field, in such a way that the detectable mass has an unaviodable stochastic behavior. Additionally, it was proved that the inherited stochasticity of the energy--momentum of the Klein--Gordon field implies decoherence for particles moving in the region where this tensor is present.
\bigskip
\bigskip

\Large{\bf Acknowledgments.}\normalsize
\bigskip

The author would like to thank A. A. Cuevas--Sosa for his 
help. 

\end{document}